\documentclass[11pt]{article} 
\usepackage[utf8]{inputenc} 
\usepackage{geometry} 
\usepackage{graphicx} 

\usepackage{booktabs} 
\usepackage{array} 
\usepackage{paralist} 
\usepackage{verbatim} 
\usepackage{subfig} 
\usepackage{color}

\usepackage{fancyhdr} 
\usepackage{sectsty}
\allsectionsfont{\sffamily\mdseries\upshape} 

\usepackage[nottoc,notlof,notlot]{tocbibind} 
\usepackage[titles,subfigure]{tocloft} 

\title{On the Statistical Foundations of Kaluza's Magnetohydrodynamics}
\author{A. R. Sagaceta-Mejía$^{1}$ and A. Sandoval-Villalbazo$^{1}$ 
\bigskip
\\ $^{1}$ Departamento de Física y Matemáticas \\  Universidad Iberoamericana \\
Prolongación Paseo de la Reforma 880, Lomas de Santa Fe CP 01219,\\ Ciudad de México.\\
 e-mail:  alma.sagaceta@correo.uia.mx, alfredo.sandoval@ibero.mx
} 

\begin{document}
	\maketitle 

	\begin{abstract}
		The introduction of electromagnetic fields into the Boltzmann equation following a 5D general relativistic approach is considered in order to establish the transport equations for dilute charged fluids in the presence of a weak electromagnetic field. The conserved 5D stress-energy tensor is evaluated using the J\"uttner function for non-degenerate relativistic gases in local equilibrium, and the evolution equations for the local thermodynamic variables are established by means of relativistic kinetic theory. An outline of the  possibilities offered by  the Kaluza-type approach to MHD is also included.
	\end{abstract}
	
	\section{Introduction}
	It is well-known that relativistic kinetic theory is necessary to describe transport processes in high temperature gases. As the parameter $z=k T/m c^2$ increases, thermal relativistic effects become more significant.  The use of the J\"utnner function:
	\begin{equation}
	f^{(0)}=\frac{n}{4 \pi c^3 z K_2(1/z)}e^{\frac{m V_\mu U^\mu}{kT}}
	\label{eq:1}
	\end{equation}
	allows to establish straightforwardly the relativistic generalization of the transport equations in the Euler regime. In Equation (\ref{eq:1})\, $n$ is the numerical density of particles,  $V_\mu$ the individual particle velocity, $U^\mu$ the hydrodynamic velocity and  $K_n(1/z)$ the modified  Bessel function  of the second type of  order $n$.
	\\
	Recently, it has been noticed that the effects of electromagnetic fields can be introduced into the Boltzmann equation using a five-dimensional space-time, according to the tenets of the general theory of relativity \cite{1}. Back in 1921, T. Kaluza showed that the  motion of individual particles in an electromagnetic field corresponds to geodesics identifying the electromagnetic field tensor with the geometric Christoffel symbols present in general relativity  using the relation 
	\begin{equation}
	F^{\mu}_{\nu}=-\frac{2}{\xi}\Gamma^{\mu}_{5 \nu}\label{eq:0}
	\end{equation}
	and proposing that the charge-mass ratio of the individual particles is related to a fifth component of the velocity through the expression  $v^{5}=\frac{q}{\xi m}$,  where $\xi=\sqrt{\frac{16\pi G \epsilon_0}{c^2}}$ is a coupling constant.
	Kaluza also showed that Maxwell's equations are contained in the Einstein field equations using the extra dimension ansatz and the \textit{cylindrical condition} $\frac{\partial}{\partial x^5 } =0 $, which corresponds to symmetry with respect to any change in the fifth coordinate \cite{2} . Following these  ideas, it is appealing  to introduce this geometric approach to kinetic theory in order to  describe the transport processes present in dilute charged gases, reproducing well-known results from a new perspective and possibly identifying new effects mainly present in the $z>1$ regime.
	\\
	This work is intended to show how kinetic theory can be applied in order to obtain the basic transport equations for a dilute charged fluid, in accordance to the Kaluza's approach to electromagnetism. To accomplish this task we have divided this paper as follows: Section 2 includes, for clarity reasons,  a review of the Euler regime in gravitational fields using the basic features of general relativity,  section 3 is devoted to the analysis of the stress-energy tensor within the 5D formalism, finally  a discussion of the possible relevance of the ideas here presented is included in the fourth section of this work.
	
	\section{Review of the Euler regime of a dilute relativistic fluid in the presence of gravity.}
	The starting point of this section is the conserved stress-energy tensor in a four-dimensional space-time, which reads
	\begin{equation}
	T^{\mu \nu}=m \int f V^{\mu} V^{\nu} d^{*}V \label{eq:2}
	\end{equation}
	where $f$ is the one-particle distribution function for individual atoms of mass $m$ evolving in space-time with four-velocity $V^{\mu} $.
	The basic transport equations are included in the expressions:
	\begin{eqnarray}
	\left(n U^\mu\right)_{;\mu}&=&0\label{eq:2.1}\\
	T^{\mu \nu}_{;\nu}&=&0\label{eq:2.2}
	\end{eqnarray}
	
	\noindent which can be obtained multiplying the relativistic Boltzmann equation by the collisional invariants ($1$, $m V^{\nu}$) and upon integration in the velocity space using the invariant velocity element $ d^{*}V $ \cite{3}. In these equations the semicolon is used to indicate a covariant derivative which  includes the effects of the gravitational field in terms of space-time curvature.
	In order to readily evaluate the stress-energy tensor in local equilibrium, it is useful to split the four-velocities present in Equation (\ref{eq:2}) in its systematic and peculiar velocities $K^\mu$. For this purpose we use Eckart's decomposition \cite{4}:
	\begin{equation}
	V^\mu=\gamma U^\mu+R^{\mu}_\nu K^{\nu}\label{eq:3}
	\end{equation}
	where the tensor $R^{\mu}_\nu$ corresponds to the product of the Lorentz transform in the comoving frame (using the hydrodynamic velocity),  and the spatial projector for a given metric $g^{\mu \nu}$ reads $h^{\mu \nu}=g^{\mu \nu}+\frac{1}{c^2}U^\mu U^\nu$ \cite{5}. In Equation (\ref{eq:3}) $\gamma$ is the usual Lorentz factor referred to the peculiar velocities. \\
	Substitution of Equation (\ref{eq:1}) and (\ref{eq:3}) into Equation (\ref{eq:2}) yields:
	\begin{equation}
	T^{\mu \nu}=\frac{n \epsilon}{c^2} U^\mu U^\nu + p h^{\mu \nu}\label{eq:76}
	\end{equation}
	where the local pressure and the internal energy are respectively given by:
	\begin{eqnarray}
	p&=&n k T \label{eq:77}\\
	\epsilon&=&m c^2 \left(\frac{K_3(1/z)}{K_2(1/z)}-z\right)\label{eq:78}
	\end{eqnarray}
	The  relativistic Euler regime is straightforwardly obtained from Equation (\ref{eq:2.1}) to (\ref{eq:2.2}),  and Equation (\ref{eq:76}) to (\ref{eq:78}). This set of equations can be simplified in terms of total time derivatives namely,
	
	\begin{eqnarray}
	\dot{n}+n U^\mu_{;\mu}&=&0 \label{eq:3.1}\\
	\tilde{\rho}\dot{U}^\nu&=&-h^{\nu\alpha}p_{,\alpha}\label{eq:3.2}
	\end{eqnarray}
	
	\noindent In Equation (\ref{eq:3.2}), \, $\tilde{\rho}=\frac{n\epsilon}{mc^2}+\frac{p}{c^2}$ is an effective density which in the non-relativistic limit $(z<<1)$ simply becomes the ordinary mass density $\rho=nm$.
	
	\noindent The system (\ref{eq:3.1}-\ref{eq:3.2}) is well-known and can be found in classical references \cite{5-0,5-1}. 
	In the next section we will apply an analogue technique in order to obtain the MHD equations in the Euler regime with the aforementioned 5D space-time.
	
	\section{Momentum balance equation in GR: the ``hidden force" and the 5D approach}

	The effects of ``external forces"  are indeed present in Equation (\ref{eq:3.2}), this is an important conceptual difference between the Newtonian and the geometric approaches.  
	In general relativity the absolute derivative $\dot{U}^\nu=\frac{\partial U^\nu}{\partial x^\alpha }U^\alpha+\Gamma^\nu_{\alpha\beta}U^\alpha U^\beta$ contains the field effects; thus Equation (\ref{eq:3.2}) can be rewritten in terms of a flat space-time total proper-time derivative as:
	
	\begin{equation}
	U^\alpha \frac{\partial U^\nu}{\partial x^\alpha }=\frac{d U^\nu}{d \tau}=-\Gamma^\nu_{\alpha\beta }U^\alpha U^\beta-\frac{1}{\tilde{\rho}}h^{\nu\alpha}p_{,\alpha}\label{Eq:122}
	\end{equation}
	
	\noindent In Equation (\ref{Eq:122}) $\Gamma^\nu_{\alpha\beta} $ is the Christoffel symbol, which contains the gravitational sources and is given by 
	
	\begin{equation}
	\Gamma^\nu_{\alpha\beta}=\frac{1}{2}g^{\nu\mu}\left(\frac{\partial g_{\beta \mu}}{\partial x^\alpha}+\frac{\partial g_{\mu \alpha}}{\partial x^\beta}-\frac{\partial g_{\alpha\beta}}{\partial x^\mu}\right) . \label{eq:4}
	\end{equation}
	
	\noindent In the simplest case of a Newtonian metric,  and for $z<<1$, Equation (\ref{eq:4}) reduces to the ordinary Euler equation in the presence of a gravitational field
	\begin{eqnarray}
	\frac{d\vec{u}}{dt}&=&-\nabla \phi-\frac{\nabla p}{\rho}\\
	\vec{g}&=&-\nabla \phi
	\end{eqnarray}
	
	In the same spirit, an analogue treatment of the Boltzmann equation can be performed in the Kaluza 5D space-time [1]. The conserved quantity $T^{AB}$ contains the effects of the electromagnetic field:

	\begin{equation}
	T^{A B}=m \int f V^{A} V^{B} d^{*}V, \label{eq:5}
	\end{equation}
	
	\noindent in Equation (\ref{eq:5}), the indices run from 1 to 5, and according to Kaluza's ansatz $(V^5=\frac{q}{\xi m})$ the new components correspond to the electric charge density and currents, so that
	\begin{equation}
	T^{5\nu}=\left[
	\begin{array}{c}
	\frac{q}{\xi}n U^\ell\\
	\frac{qnc}{\xi}
	\end{array}
	\right]
	\label{eq:6}
	\end{equation}
	
	\noindent the component $T^{55}$ in Equation (\ref{eq:5}) is not present in the balance equations due to the cylindrical condition, so it is not necessary to show its explicit computation. \\
	It is convenient to discuss carefully both the continuity equation and the  momentum balance in this geometrical picture. For the particle flux, direct integration of the Boltzmann equation yields:
	
	\begin{equation}
	N^A_{;A}=0 
	\end{equation}
	where $N^A=\int f V^A d^* V$.
	
	Using the cylindrical condition, neglecting second order terms in the field and considering the fact that the electromagnetic tensor is antisymmetric, we obtain no changes in the ordinary  general relativistic continuity equation
	
	\begin{equation}
	\frac{\partial N^A}{\partial x^A}+\Gamma^A_{AB}N^B=\frac{\partial N^\nu}{\partial x^\nu}+\Gamma^\nu_{\nu \mu}N^\mu=0
	\end{equation}
	\noindent This equation is consistent with  charge conservation  $T^{5\nu}_{;\nu}=0$, by means of the density-flux expression (\ref{eq:6}), so that charge and particle  balances are identical. In the derivation of Equation (\ref{eq:6}) use has been made of the metric tensor 
	
	\begin{equation}
	g_{AB}=\left[
	\begin{array}{ccccc}
	1&0&0&0&A^1 \xi\\
	0&1&0&0&A^2 \xi\\
	0&0&1&0&A^3 \xi\\
	0&0&0&-1&\frac{\phi}{c}\xi\\
	A^1 \xi&A^2 \xi&A^3 \xi&\frac{\phi}{c}\xi&1\\
	\end{array}
	\right]
	\end{equation}
	and the standard expressions
	\begin{eqnarray}
	F^{\mu\nu}&=&A^{\nu,\mu}-A^{\mu,\nu}\\ \label{eq:0.1}
	A^\mu&=&\left[\vec{A},\frac{1}{c}\phi\right]\\ \label{eq:0.2}
	\vec{E}&=&-\nabla \phi\\ \label{eq:0.3}
	\vec{B}&=&\nabla\times\vec{A} \label{eq:0.4}
	\end{eqnarray}

	\noindent Equation (\ref{eq:4}) and Equation (\ref{eq:0.1}) to (\ref{eq:0.4}) are consistent up to first order with (\ref{eq:0}). \\ For the momentum balance it is convenient to introduce the covariant derivative of a second rank tensor given by
	
	\begin{equation}
	T^{AB}_{;B}=\frac{\partial T^{AB}}{\partial x^B}+\Gamma^A_{BL}T^{BL}+\Gamma^B_{BL}T^{AL}=0
	\end{equation}
	
	\noindent Using the cylindrical condition and expanding the sums in order to separate the electromagnetic contributions to the balance equation, one obtains for the ordinary space-time components:
	
	\begin{equation}
	T^{\nu B}_{;B}=\frac{\partial T^{\nu\beta}}{\partial x^\beta}+\Gamma^\nu_{\beta\lambda}T^{\beta\lambda}+\Gamma^\beta_{\beta\lambda}T^{\nu\lambda}+\Gamma^\nu_{\beta 5}T^{\beta 5}+\Gamma^\nu_{5\lambda}T^{\beta 5}=0
	\label{eq:7}
	\end{equation}
	
	\noindent  The four-dimensional Euler equation can be identified with the  first three terms of Equation (\ref{eq:7}). The last two terms correspond to the effects of the electromagnetic field. The result thus obtained is:
	
	\begin{equation}
	\tilde{\rho}\dot{U}^\nu+h^{\nu\alpha}p_{,\alpha}=-2\Gamma^\nu_{\lambda 5}T^{5\lambda}\label{eq:8}
	\end{equation}
	
	\noindent The Christoffel symbol included in Equation (\ref{eq:8}) is easily identified with the electromagnetic field tensor in accordance to Equation (\ref{eq:0}), so that the basic magnetohydrodynamic momentum balance is recovered:
	
	\begin{equation}
	\tilde{\rho}\dot{U}^\nu+h^{\nu\alpha}p_{,\alpha}=F^\nu_{\lambda}J^{\lambda}=\left[
	\begin{array}{c}
	q n \vec{E}+\vec{J}\times \vec{B}\\
	\vec{J}\cdot\vec{E}
	\end{array}
	\right]
	\end{equation}
	
	\noindent where $J^\nu$ is the current density four-vector given by
	\begin{equation}
	J^\nu=n q \left[
	\begin{array}{c}
	U^\ell\\
	c
	\end{array}
	\right] \end{equation}
	
	\noindent and $\vec{E},\vec{B}$ correspond to the electric and magnetic fields respectively.

	In Equation (\ref{eq:8}) relativistic effects arise, not only from the space-time  curvature approach, but also from the thermodynamic effects included in $\tilde{\rho}$. It is interesting to notice that this equation corresponds to that reported using a flat space-time and the ordinary external force approach in the context of special relativity \cite{6,6-0}.
	
	The derivation here presented is novel and broadens the ideas earlier exposed in Ref. \cite{1}.
	\section{Final remarks}
	
	
	Space-time curvature has been successfully introduced in the Boltzmann equation for the case of gravitational fields since the 1960's \cite{7.0}.
	Interesting results regarding gravitational collapse (Jeans instability)  for dissipative fluids and bulk viscosity are now well understood  thanks to relativistic kinetic theory \cite{7.01}. 
	The use of GR can also be applied to the case of the electromagnetic field by means of the Kaluza's approach to field theory. The first work that included dissipative processes in a Kaluza-type MHD was performed in a phenomenological fashion sixteen years ago \cite{7}. Later on, an attempt to establish a wave equation for heat conduction in a simple charged fluid using these techniques was also proposed phenomenologically \cite{8}. Nevertheless, only recently the kinetic foundations were analyzed through the Boltzmann equation \cite{1}.
	Success has been achieved in the case of the Euler charged fluid and work is in progress in the Navier-Stokes regime for single component gases. \\
	
	A direct statistical treatment of dissipative effects in Kaluza's MHD involves the use of the Chapman-Enskog expansion in which the distribution function up to first order in the gradients has the form: 
	
	\begin{equation}
	f=f^{(0)}+f^{(1)}
	\end{equation}
	\noindent where $f^{(0)}$ is given by Equation (\ref{eq:1}) and $f^{(1)}$ has the approximate form \cite{1}
	
	\begin{equation}
	f^{(1)}=-\tau V^A \left(\frac{\partial f^{(0)}}{\partial n}n_{,A}+\frac{\partial f^{(0)}}{\partial T}T_{,A}+{\frac{\partial f^{(0)}}{\partial U^B} \label{eq:9} U^B_{;A}}\right)
	\end{equation}
	
	\noindent In Equation (\ref{eq:9}) $\tau$ corresponds to the relaxation parameter of the BGK approximation and the electromagnetic effects are contained in the covariant derivatives taken over the complete 5D space-time. Our approach allows to identify thermal-electric and thermal-magnetic effects as being caused by space-time curvature properties of space-time which in contrast are  viewed as external forces in Newtonian physics.\\
	
	Future work includes a generalization to the Burnett regime, the use  of a more accurate model for the Kernel in the Boltzmann equation and the careful reproduction of cross-effects (Soret, Dufour) for binary systems. The mathematical methods of general relativity have already shown its potentialities in astrophysical problems regarding rarefied and low density fluids in the presence of high curvature and with  extremely high values of the parameter $z$. It is interesting to notice that most of these GR procedures can in principle be introduced to the study of plasma physics using the ideas proposed in the present work. Thus, it is possible that this approach will provide new insights for important issues  such as causality and stability in these thermodynamic systems.

	\section{Acknowledgements}
	The authors acknowledge support from CONACyT (M\'exico) through grant number CB2011/167563. The authors also wish to thank A.L. Garc\'ia-Perciante for her valuable comments to this work.

	

\end{document}